\documentclass[prc,superscriptaddress,showpacs,amsmath,amssymb,twocolumn]{revtex4-1}
\usepackage{amsmath, amsthm, amssymb, braket}

\usepackage{amsfonts}

\usepackage{graphicx}
\usepackage{dcolumn}
\usepackage{bm}

\usepackage{color}

\newcommand{\be}{\begin{equation}}
\newcommand{\ee}{\end{equation}}

\newcommand{\br}{\boldsymbol{r}}
\usepackage{ulem}

\begin{document}

\title{Charge-exchange dipole excitations in neutron-rich nuclei: 
$-1 \hbar \omega_0$, anti-analog pygmy, and anti-analog giant resonances
}

\author{Kenichi Yoshida}
\affiliation{Department of Physics, Kyoto University, Kyoto, 606-8502, Japan}

\date{\today}

\begin{abstract}
The occurrence of the low-lying charge-exchange non spin-flip dipole 
modes below the giant resonance in neutron-rich nuclei 
is predicted on the basis of nuclear density functional theory. 
The ground and excited states are described in the framework of the 
self-consistent Hartree-Fock-Bogoliubov and the proton-neutron 
quasiparticle-random-phase approximation employing 
a Skyrme-type energy density functional. 
The model calculations are performed for the spherical neutron-rich Ca, Ni, and Sn isotopes. 
It is found that 
the low-lying states appear sensitively to the shell structure 
associated with the $-1 \hbar \omega_0$ excitation 
below the Gamow-Teller states. 
Furthermore, the pygmy resonance emerges below the giant resonance 
when the neutrons occupy the low-$\ell (\ell \leq 2 -3)$ orbitals 
analogous to the pygmy resonance seen in the electric dipole response. 
\end{abstract}

\maketitle


Response of atomic nuclei to external fields 
reveals the occurrence of a variety of modes of excitation, 
and study for nuclei far from the $\beta$-stability line 
has been a major subject of nuclear physics. 
The type of nuclear response is characterized by the transferred 
angular momentum $L$, spin $S$, and isospin $T$~\cite{har01}, 
and the combination of these quantum numbers makes the excitation modes rich in nuclear systems.
Among various types of excitation, the isovector (IV) dipole $(T=1, L=1)$ resonance 
has been investigated for many nuclei throughout the periodic table over a long period of time 
as a central issue in the photo-nuclear process~\cite{BF75}.  
The out-of-phase vibration of neutrons and protons, 
known as the giant dipole resonance (GDR), 
is the most established collective mode in nuclear system~\cite{har01}. 
Recently, a number of works have been devoted to
the quest for exotic modes of excitation in neutron-rich nuclei. 
A representative example is the low-energy dipole mode, or 
the pygmy dipole resonance (PDR)\cite{paa07, aum13}. 
In view of 
the correlation between the strength distribution of the PDR and 
the neutron skin thickness, 
the equation of state of neutron matter 
has been extensively investigated~\cite{pie06, tso08, car10, ina11, ina13}. 
For the charge-exchange mode of excitation, 
the appearance of low-energy Gamow-Teller states in the light neutron-rich nuclei 
and the mechanism for the fast $\beta$ decay
have been discussed in Ref.~\cite{sag93}.

The IV dipole resonance 
was also observed by the charge-exchange ($p,n$) reaction 
as an anti-analog state of the GDR (AGDR)~\cite{bai80, ste80}, 
though it was difficult to distinguish the non spin-flip component from  
the spin-flip axial-vector dipole resonance (SDR) 
because both the $S=0$ and $S=1$ components 
can be excited by the hadronic ($p,n$) reaction~\cite{ost81, nis85,aus01}. 
Investigation of the IV excitations 
not only in the $T_z=0$ excitation but also in the $T_z=\pm 1$ excitations 
could lead us to a comprehensive understanding 
of the nature of the IV modes of excitation, 
such as the isospin character of the PDR.
Furthermore, the IV dipole responses as well as 
the Fermi and Gamow-Teller ($L=0$) responses 
play important roles 
for the description of the nuclear responses associated with 
the low-energy neutrinos and the low-energy weak processes~\cite{eji00}.

Since the authors of Ref.~\cite{kra13} proposed 
a new approach to put a constraint on the neutron skin thickness 
based on the excitation energy of the AGDR, 
the charge-exchange non spin-flip dipole resonance has again attracted interests. 
Quite recently, separation of the AGDR from the SDR 
was successfully achieved by using the polarized proton beam for the $^{208}$Pb $(p,n)$ reaction~\cite{yas13}. 
And a detailed theoretical analysis was made on the correlation between 
the neutron skin thickness and 
the excitation energy of the AGDR with resect to the isobaric analog state (IAS)~\cite{cao15}.

\begin{figure}[b]
\begin{center}
\includegraphics[scale=0.45]{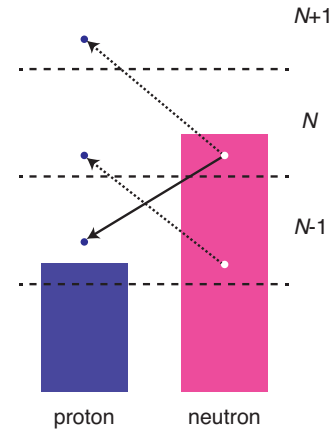}
\caption{(Color online)
Schematic picture for the charge-exchange 
particle-hole excitations with 
negative parity in neutron-rich nuclei, 
where the Fermi levels of neutrons and protons are located in a different major harmonic-oscillator shell of $N$ and $N-1$. 
The solid and dotted arrows correspond to the $-1 \hbar \omega_0$ and $1 \hbar \omega_0$ excitations, respectively. 
The particle continuum threshold is located either in the $N$ shell (near the neutron drip line ) or in the $N+1$ shell. }
\label{ffv}
\end{center}
\end{figure}

\begin{figure*}[t]
\begin{center}
\includegraphics[scale=0.22]{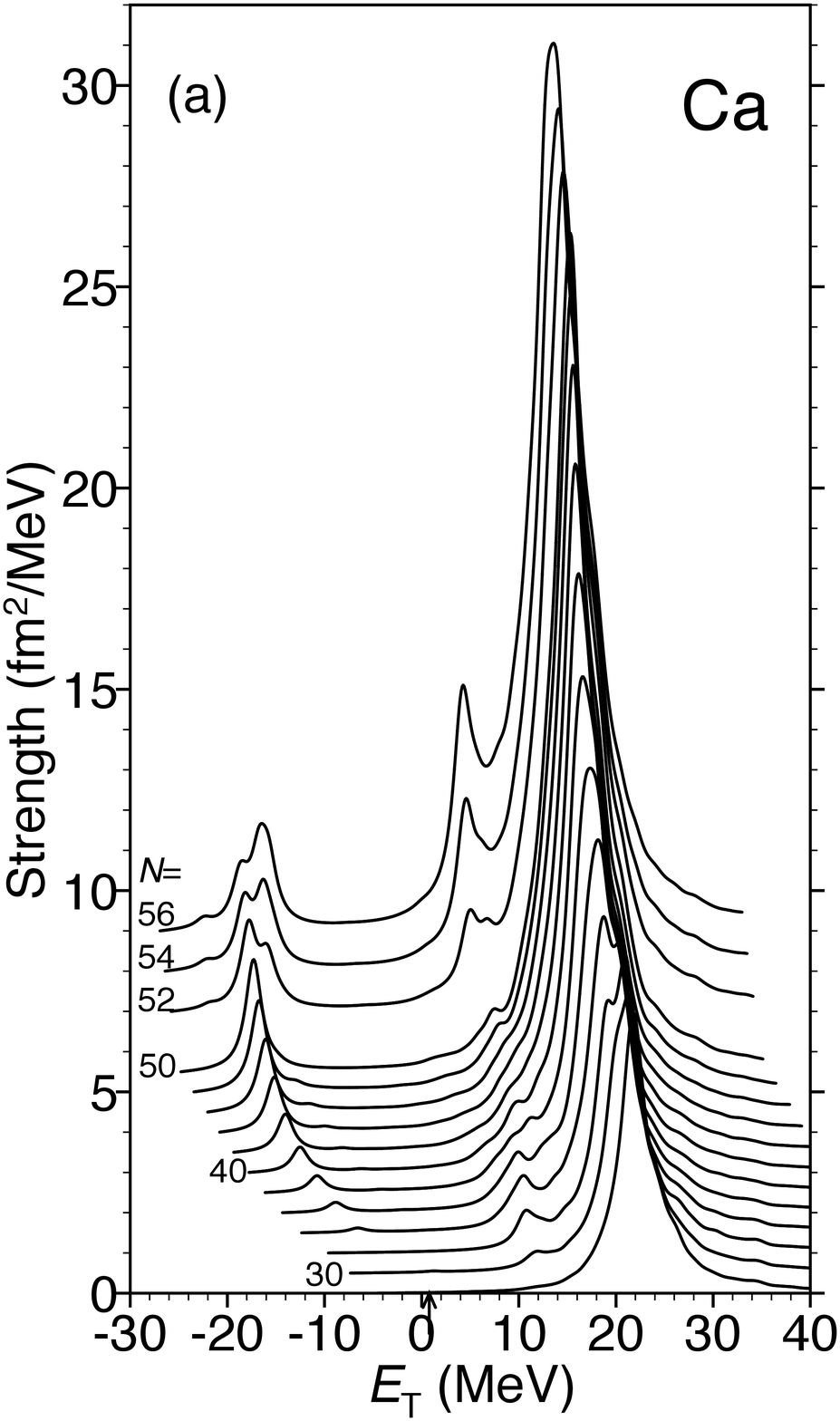}
\includegraphics[scale=0.22]{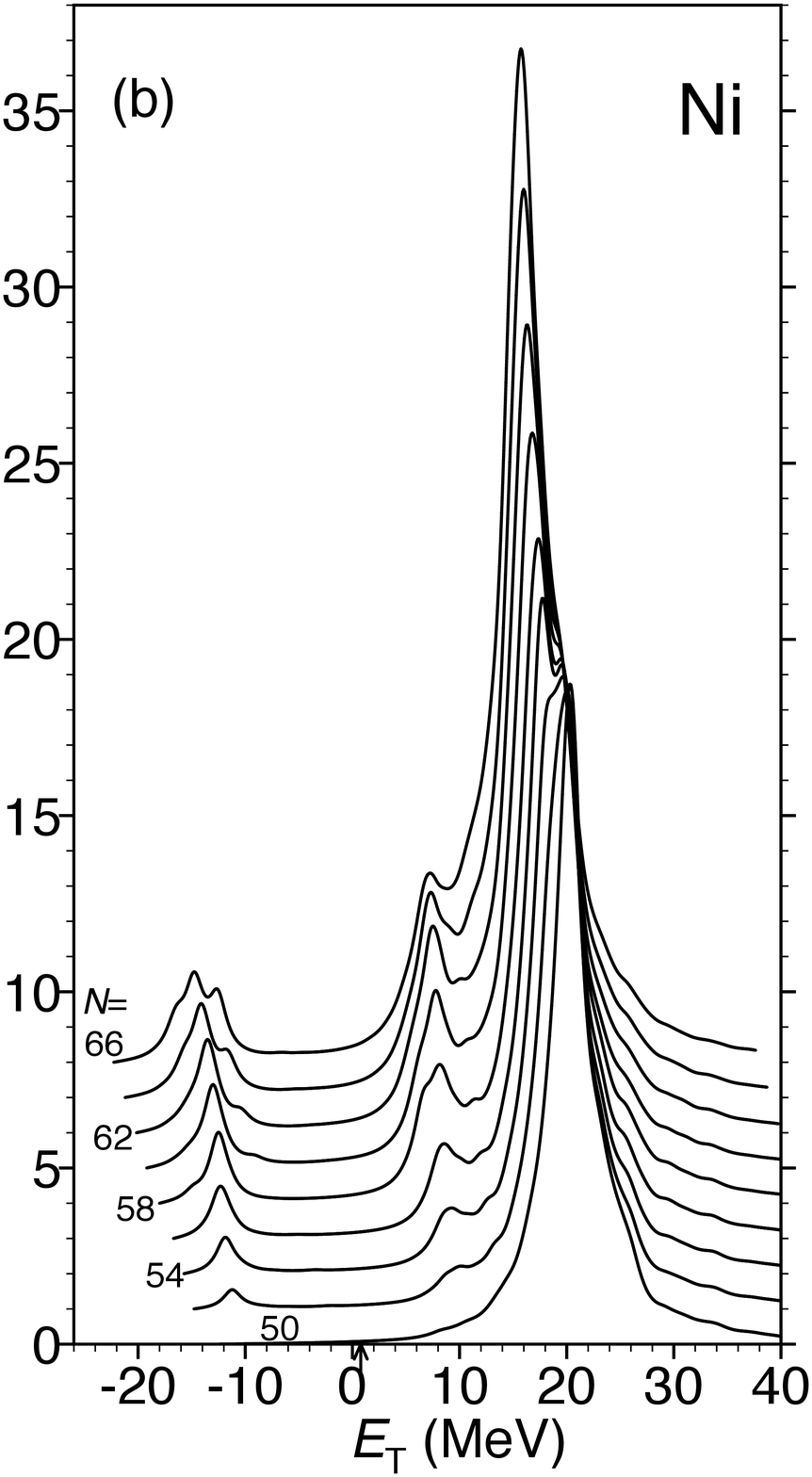}
\includegraphics[scale=0.22]{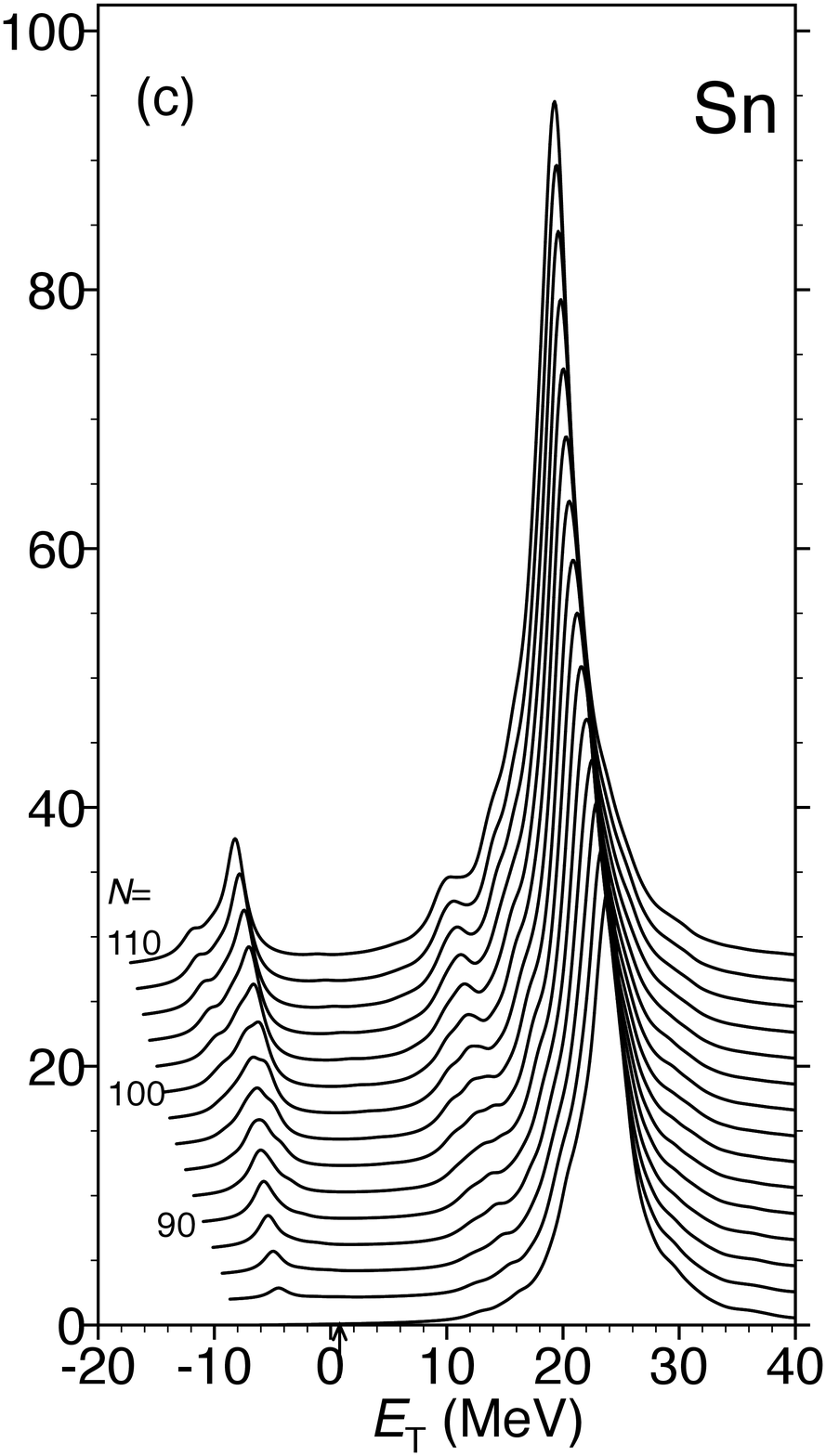}
\caption{
Charge-exhange ($T_z = -1$) dipole strength distributions (shifted) for the neutron-rich Ca, Ni, and Sn isotopes 
as functions of the excitation energy with respect to the ground state of the target nuclei employing the SkM* functional.
The width $\gamma=2.0$ MeV is included for smearing the distributions. The arrow indicates the threshold for the $\beta$ decay.}
\label{strength}
\end{center}
\end{figure*}

In the present article, I investigate 
the charge-exchange non spin-flip vector dipole ($T=1, T_z=-1, L=1, S=0$) response. 
I consider the neutron-rich systems 
where the Fermi levels of neutrons and protons are apart by one major harmonic-oscillator shell 
as shown in Fig.~\ref{ffv}. 
In such a situation, 
the excitation of a neutron-hole in the $N$-shell to a proton-particle in the $N-1$-shell 
is available uniquely at low energy of $-1 \hbar \omega_0$. 
The neutron-rich $^{60}$Ca, $^{78}$Ni and $^{132}$Sn 
nuclides are located at the boundary of this situation, 
and the roles of neutron excess on the low-lying states 
in these isotopes are studied. 
The appearance of the low-lying dipole states 
below the Gamow-Teller states corresponding to the $0 \hbar \omega_0$ mode 
may have a strong impact on the $\beta$-decay rate and 
the $\beta$-delayed neutron emission probability. 
The effect of the first-forbidden transitions on the $\beta$-decay rate have been investigated around 
the $r$-process waiting-point nuclei~\cite{bor03}, where 
the contribution of both the vector and axial-vector dipole states is present.

In a framework of the nuclear energy-density functional (EDF) method I employed, 
the excited states $| i \rangle$ of the daughter nucleus are described as 
a one-phonon excitation built on the ground state $|0\rangle$ of the mother (target) nucleus:
\begin{align}
| i \rangle &= \hat{\Gamma}^\dagger_i |0 \rangle, \\
\hat{\Gamma}^\dagger_i &= \sum_{\alpha \beta}\left\{
X_{\alpha \beta}^i \hat{a}^\dagger_{\alpha,\nu}\hat{a}^\dagger_{\beta, \pi}
-Y_{\alpha \beta}^i \hat{a}_{\bar{\beta},\pi}\hat{a}_{\bar{\alpha},\nu}\right\},
\end{align}
where $\hat{a}^\dagger_\nu (\hat{a}^\dagger_\pi)$ and  $\hat{a}_\nu (\hat{a}_\pi)$ are 
the neutron (proton) quasiparticle creation and annihilation operators. 
The quasiparticles (qp's) $\alpha, \beta$ were obtained as a self-consistent solution 
of the Hartree-Fock-Bogoliubov (HFB) equation~\cite{dob84, bul80}. 
To describe the developed neutron skin and the neutrons pair correlation 
coupled with the continuum states, 
I solved the HFB equations in the coordinate space using cylindrical coordinates
$\boldsymbol{r}=(\rho,z,\phi)$ with a mesh size of
$\Delta\rho=\Delta z=0.6$ fm and a box
boundary condition at $(\rho_{\mathrm{max}},z_{\mathrm{max}})=(14.7, 14.4)$ fm.
The qp states were truncated according to the qp 
energy cutoff at 60 MeV, and 
the qp states up to the magnetic quantum number $\Omega=23/2$
with positive and negative parities were included. 
The phonon states, the amplitudes $X^i, Y^i$ and the vibrational frequency $\omega_i$, 
were obtained in the proton-neutron quasiparticle random-phase approximation (pnQRPA).
The two-body interaction for the pnQRPA equation was derived self-consistently from the EDF. 
I introduced the truncation for the two-quasiparticle (2qp) configurations in the QRPA calculation,
in terms of the 2qp-energy as 60 MeV. 
More details of the calculation scheme are given in Ref.~\cite{yos13}.

For the normal (particle-hole) part of the EDF,
I employed the SkM* functional~\cite{bar82}. 
For the pairing energy, I adopted the one in Ref.~\cite{yam09}
that depends on both
the isoscalar and isovector densities, 
in addition to the pairing density, with the parameters given in
Table~III of Ref.~\cite{yam09}. 
The same pairing EDF was employed for the pn-pairing interaction 
in the pnQRPA calculation, 
while the linear term in the isovector density was dropped.
The spin-triplet pairing interaction, the presence of which is controversial, 
was not included in the present calculation for simplicity.

In Fig.~\ref{strength}, the transition-strength distributions for the dipole operator 
are presented as functions of the 
excitation energy $E_{\rm T}$ with respect to 
the ground state of the mother (target) nucleus: 
\begin{align}
S (E_{\rm T})
&=\sum_{K}\sum_{i} \dfrac{\gamma/2}{\pi}\dfrac{ R_i ^2}
{[E_{\rm T}- \{ \hbar\omega_{i} - ( \lambda_\nu - \lambda_\pi)\}]^{2}+\gamma^{2}/4}, \label{response} \\
R_i &=  \langle i | \hat{F}_{1K}^- |0 \rangle   = \langle 0 |[\hat{\Gamma}_i, \hat{F}^-_{1K}] |0 \rangle
= \sum_{\alpha \beta} M^i_{\alpha \beta},
\label{mat_ele}
\end{align} 
where $\lambda_\nu (\lambda_\pi)$ is the chemical potential for neutrons (protons) and the dipole operator is defied as
\begin{equation}
\hat{F}^-_{1K} = \sum_{\sigma}\int d\boldsymbol{r} 
rY_{1K}(\hat{r}) \hat{\psi}^\dagger_\pi(\boldsymbol{r}  \sigma)  \hat{\psi}_\nu(\boldsymbol{r}  \sigma) \label{dipole_op}
\end{equation}
in terms of the nucleon field operators. 
The strength distributions for the QRPA energy $\hbar \omega \geq 0$ MeV are presented. 
It is noted that I found the ground state of the neutron-rich Sn isotopes beyond $N=96$ is deformed in the present calculation, 
but I show in the present article the results obtained in the spherically constrained calculation 
to avoid the discussion on the nuclear deformation. 
It would be interesting to investigate the deformation effect as a future work.

\begin{figure}[t]
\begin{center}
\includegraphics[scale=0.22]{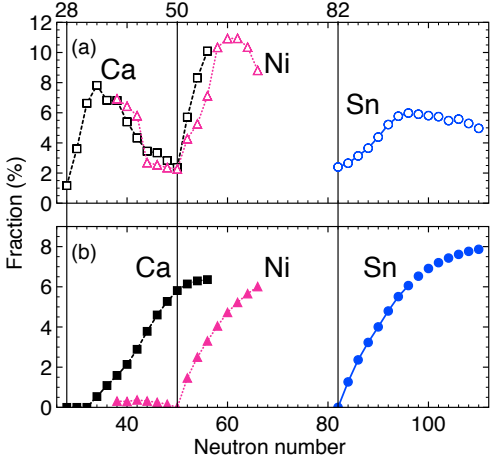}
\caption{(Color online)
(a) Fraction of the summed strengths of the pygmy dipole resonance 
to the total sum of strengths. 
(b) Fraction of the summed strengths of the $-1 \hbar \omega_0$ states to the total sum of strengths.}
\label{fraction}
\end{center}
\end{figure}

One can see a low-energy peak below $E_{\rm T}=0$ MeV and 
its development depends on the neutron number.
To see clearly the neutron number dependence of the development of 
the low-energy states, 
I show in Fig.~\ref{fraction}(b) the fraction of the transition strengths to these states 
to the total sum of the calculated strengths for the dipole operator (\ref{dipole_op}). 
Here, the low-energy strength associated with the $-1 \hbar \omega_0$ excitation was evaluated by 
the summed strengths below the QRPA energy of $\hbar \omega=15$ MeV, where one has only the low-energy $-1 \hbar \omega_0$ states. 
As mentioned in the introduction, one can expect the appearance of the $-1 \hbar \omega_0$ excitation mode 
beyond $N=40$ for the Ca isotopes due to the $\pi 1f_{7/2} \otimes \nu 1g_{9/2}$ excitation. 
In the present calculation, I found the partial occupation of the $1g_{9/2}$ orbital above $N=34$
due to the pair correlation of neutrons. Thus, the low-energy strength develops gradually starting from $N=34$ to $N=50$. 
Figure~\ref{strength}(a) shows that the low-energy peak consists of several states beyond $N=50$. 
Indeed, the $2d_{5/2}$ orbital appears as a quasi-neutron resonance at low energy, 
and the $\pi 1f_{5/2}\otimes \nu 2d_{5/2}$ excitation begins to participate in generating the low-energy peak 
besides the $\pi 1f_{7/2} \otimes \nu 1g_{9/2} $ excitation.
For the neutron-rich Ni isotopes, one can see the low-energy strength starts increasing at $N=50$. 
The low-energy states are mainly constructed by the $\pi 2p_{3/2} \otimes \nu 2d_{5/2}$ and 
$\pi 2p_{1/2} \otimes \nu 2d_{3/2}$ excitations. 
Beyond $N=58$, the neutron continuum states play a major role. 
For the neutron-rich Sn isotopes, the low-energy strength starts increasing at $N=82$ due to 
the $\pi 2d_{5/2} \otimes \nu 2f_{7/2}$ excitation. 
In the very neutron-rich isotopes with $N=100-110$, one can see 
the low-energy peak structure is well developed. 
This is because the $\pi 1h_{11/2} \otimes \nu 1i_{13/2}$ excitation plays a dominant role.
One can say the low-energy $-1 \hbar \omega_0$ states are weakly collective, 
because the combination of the particle-hole or 2qp excitations 
satisfying the selection rule ($\Delta l=1, \Delta j=1$) is limited. 
So, the excitation energy and the transition strength are sensitive to the shell structure, 
and thus to the EDF employed for the calculation.

\begin{figure}[t]
\begin{center}
\includegraphics[scale=0.2]{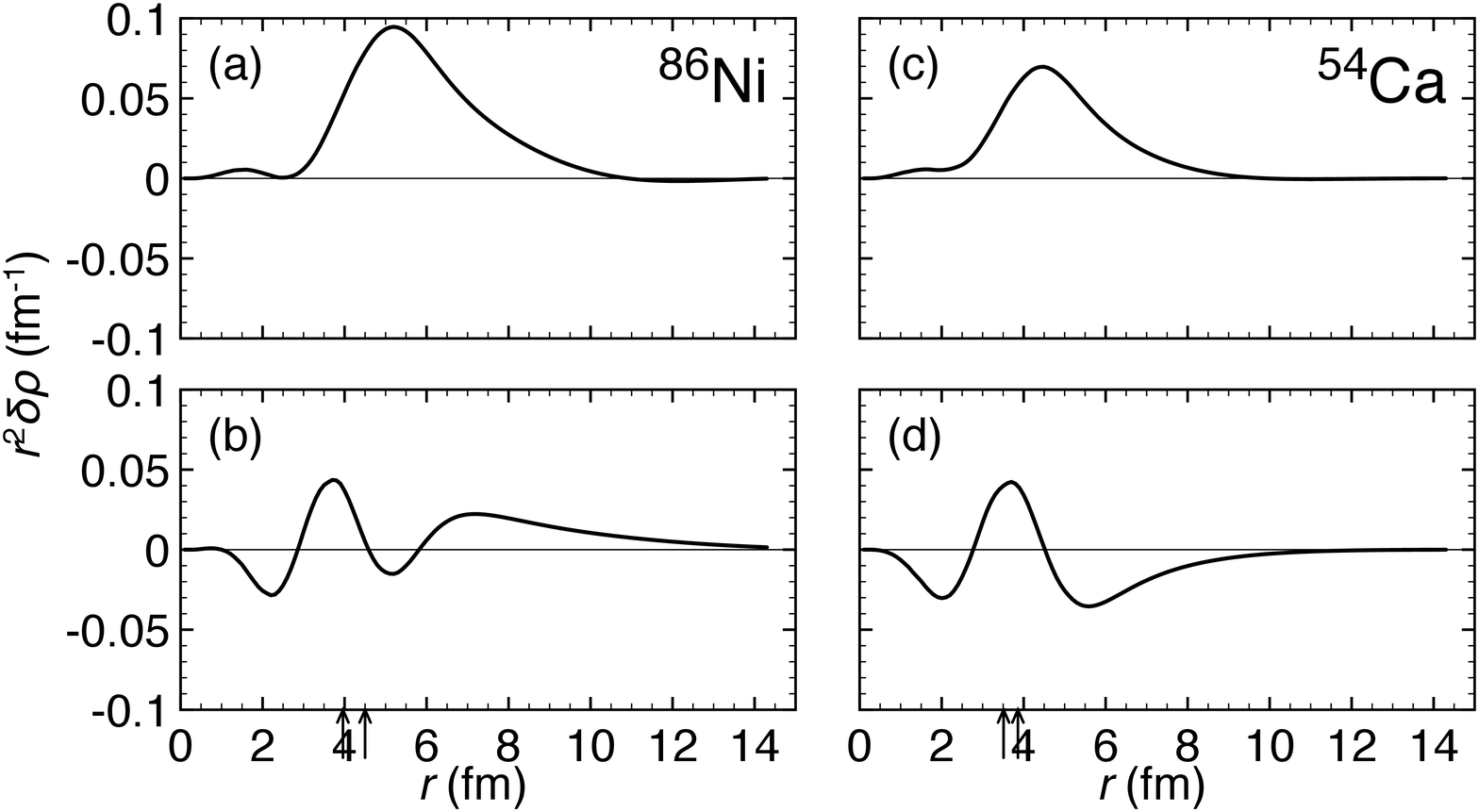}
\includegraphics[scale=0.2]{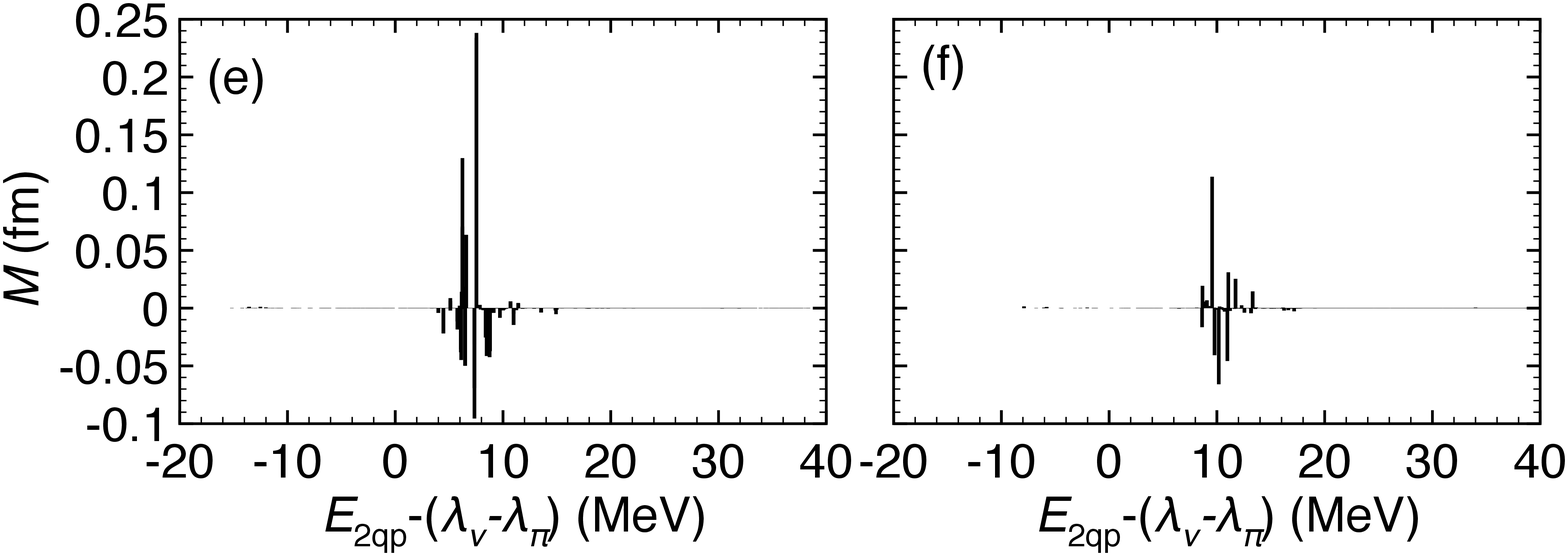}
\caption{
Transition densities to (a) the giant resonance at $\hbar \omega=35.4$ MeV ($E_{\rm{T}}=17.4$ MeV) 
and (b) the pygmy resonance at $\hbar \omega=25.8$ MeV ($E_{\rm{T}}=7.8$ MeV) in $^{86}{\rm Ni}$, 
and to (c) the state at  $\hbar \omega=31.2$ MeV ($E_{\rm{T}}=18.8$ MeV) and 
(d) the state at $\hbar \omega=23.0$ MeV ($E_{\rm{T}}=10.6$ MeV)  in $^{54}{\rm Ca}$. 
The arrows indicate the root mean square radii of neutrons and protons. 
Matrix element $M^{i}_{\alpha \beta}$ for the pygmy resonance in (e) $^{86}$Ni and (f) $^{54}$Ca 
as a function of the unperturbed excitation energy.}
\label{trans_den}
\end{center}
\end{figure}

In Fig.~\ref{strength}, one sees a prominent peak or two peaks depending on the neutron number at $E_{\rm T} = 10-20$ MeV. 
Although the isospin of the QRPA eigenstates is not a good quantum number, 
the $T_0-1$ component is dominant in the neutron rich nuclei~\cite{krm83}, with $T_0$ being the isospin 
of the ground-state of the mother nucleus. 
Therefore, the resonance peak around 20 MeV corresponds to the AGDR. 
Just below the AGDR peak energy, the concentration of the strengths or the shoulder structure 
is developed depending on the neutron number. 
I am going to discuss what the resonance structure below the AGDR is. 
Figure~\ref{trans_den} shows the transition densities
\be
\delta \rho_i(\br) = \langle i | \sum_{\sigma} \hat{\psi}^\dagger_\pi(\br \sigma) \hat{\psi}_\nu(\br \sigma)| 0 \rangle
\ee
to the states at $E_{\rm T}=17.4$ MeV ($\hbar \omega=35.4$ MeV) and 
$E_{\rm T}=7.8$ MeV ($\hbar \omega=25.8$ MeV) in $^{86}$Ni,  
and the states at $E_{\rm T}=18.8$ MeV ($\hbar \omega=31.2$ MeV) and 
$E_{\rm T}=10.6$ MeV ($\hbar \omega=23.0$ MeV) in $^{54}$Ca as examples. 
The transition density to the high-energy state represents the pure IV character 
around and outside the nuclear surface. This confirms that the giant resonance seen 
in the charge-exchange dipole excitation corresponds to an analog of the GDR. 
This state is strongly collective for the IV dipole operator~(\ref{dipole_op}). 
On the other hand, 
a very different behavior is observed for the lower-lying resonance: 
It shows a complicated spatial structure. 
Inside the nucleus the IV matrix element vanishes, 
and the tail structure is developed far outside the nuclear surface.  
This suggests the lower-lying resonance observed below the AGDR 
corresponds to an analog of the PDR, 
where the transition density has an isoscalar (IS) character inside the nucleus 
and both the IS and IV transition densities are developed far from the nuclear surface.
Figures~\ref{trans_den}(e) and \ref{trans_den}(f) shows 
the matrix element $(\ref{mat_ele})$ for the pygmy resonance in $^{86}$Ni and $^{54}$Ca. 
Many 2qp excitations around $E_{\rm{T}}=10$ MeV take part in constructing the pygmy resonance, 
but they are incoherent for the IV dipole operator. 
And the excitation energy is unchanged even with the presence of the residual interactions.

Figure~\ref{fraction}(a) displays the isotopic dependence of the development of the pygmy resonance. 
To evaluate the strengths of the pygmy resonance, I first defined 
the mean frequency of the $1 \hbar \omega_0$ excitation by
\be
\bar{\omega}
=\dfrac{\sum_{\hbar \omega_i > 15 {\rm MeV}} \omega_i R^2_i }{\sum_{\hbar \omega_i > 15 {\rm MeV}} R^2_i },
\ee
including both the AGDR and the pygmy resonance. 
Then, the pygmy strength was evaluated by the summed strengths 
in the QRPA energy of 15 MeV $< \hbar \omega < \hbar \bar{\omega} - \Gamma$. 
Here, I set the width parameter of the giant resonance $\Gamma$ as 7 MeV to exclude the contribution of the AGDR. 
One sees a sudden jump in the fraction at $N=28 \to 30$, $50 \to 52$, 
and a slightly moderate increase at  $N=82 \to 84$. 
The neutron numbers $N=30, 52$, and 84 
correspond to the occupation of the $2p_{3/2}, 2d_{5/2}$, and $2f_{7/2}$ orbitals, respectively. 
The pygmy resonance is constructed by the p-h excitation from the loosely-bound neutron 
to the proton in the continuum states. 
The evolution of the transition strength is consistent with the findings in Refs.~\cite{ina11, eba14}, 
where the particle-hole excitations from the weakly-bound low-$\ell$ orbital 
play a decisive role for the emergence of the PDR. 
This also suggests the lower-lying resonance just below the AGDR in the charge-exchange dipole excitation 
can be considered as an anti-analog of the PDR (APDR). 
The mechanism for the appearance of the APDR is very similar to that of the PDR, 
and thus one can expect the emergence of the APDR irrespective of the EDF employed. 

\begin{figure}[t]
\begin{center}
\includegraphics[scale=0.2]{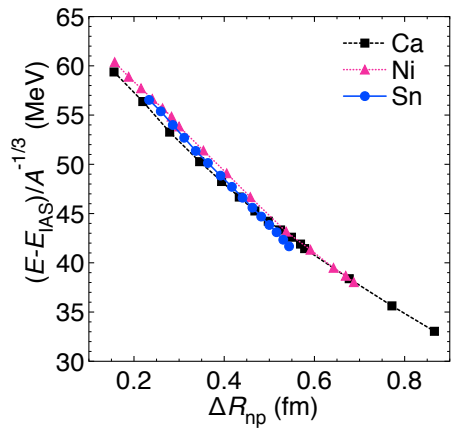}
\caption{(Color online)
Difference between the mean excitation energy of the $1 \hbar \omega_0$ states $(E=\hbar \bar{\omega})$ 
and that of the IAS $(E_{\rm IAS})$ divided by $A^{-1/3}$ 
as a function of the calculated neutron skin thickness of the Ca, Ni, and Sn isotopes.}
\label{energy}
\end{center}
\end{figure}

Finally, I am going to examine 
the correlation between the neutron skin thickness 
and the energy of charge-exchange dipole excitation modes. 
Figure~\ref{energy} shows the difference 
between the mean excitation energy $E=\hbar \bar{\omega}$ of the $1 \hbar \omega_0$ states  
and the excitation energy $E_{\rm IAS}$ of the IAS  as a 
function of the calculated neutron skin thickness of the Ca, Ni, and Sn isotopes. 
To remove the trivial mass-number dependence, 
the excitation energy is divided by $A^{-1/3}$. 
Then, I observe that the mean excitation energy of the $1 \hbar \omega_0$ states 
with respect to the IAS is negatively correlated with the neutron skin thickness. 
The magnitude of the neutron-skin thickness and the energy difference $E-E_{\rm IAS}$ 
may depend on the EDF used for the calculation. 
To extract the neutron skin thickness from the measurement, 
one needs to employ the EDF's with different symmetry energy as done in Refs.~\cite{kra13, cao15}.

In summary, 
I carried out the systematic investigation of 
the charge-exchange non spin-flip vector dipole response 
in the neutron-rich singly-closed-shell nuclei, Ca, Ni, and Sn isotopes, 
by means of the fully self-consistent pnQRPA with 
the Skyrme EDF. 
There have been discussions only on the giant resonance 
since its discovery in 1980.
I found here the emergence of the low-lying resonances 
corresponding to the $-1 \hbar \omega_0$ excitation and 
the anti-analog PDR (APDR) below the anti-analog giant dipole resonance (AGDR).
The low-energy $-1 \hbar \omega_0$ mode below the $L=0$ states 
are generated by the 2qp excitations satisfying the selection rule 
near the Fermi levels, 
thus the occurrence of it is quite sensitive to the shell structure. 
The APDR strength shows a strong enhancement 
when the neutrons occupy the low-$\ell$ orbital, 
and a moderate increase even for $\ell = 3$. 
The mean excitation energy of the $1 \hbar \omega_0$ states 
including both the APDR and AGDR with respect to the IAS 
is negatively correlated with the development of the neutron skin thickness. 

This work was supported by the JSPS KAKENHI (Grant No. 16K17687). 
The numerical calculations were performed on CRAY XC40 
at the Yukawa Institute for Theoretical Physics, Kyoto University, and 
on COMA (PACS-IX) at the Center for Computational Sciences, University of Tsukuba.

\end{document}